# Automatic laminectomy cutting plane planning based on artificial intelligence in robot-assisted laminectomy surgery

**Authors:** Zhuofu Li[1,2,3, *], M.D, Yonghong Zhang[4, *], M.S, Chengxia Wang[1,2,3, *], M.D, Shanshan Liu[1,2,3], M.D, Xiongkang Song[4], M.S, Xuquan Ji[5], Ph.D, Shuai Jiang[1,2,3], M.D, Woquan Zhong[1,2,3], M.D, Lei Hu[4, #], Ph.D, Weishi Li[1,2,3, #], M.D.

1. Department of Orthopaedics, Peking University Third Hospital, 49 North Garden Road, Haidian District, Beijing 100191, China
2. Beijing Key Laboratory of Spinal Disease Research
3. Engineering Research Center of Bone and Joint Precision Medicine, Ministry of Education
4. School of Mechanical Engineering and Automation, Beihang University, Beijing, China
5. School of Biological Science and Medical Engineering, Beihang University, Beijing, China

*Zhuofu Li, Yonghong Zhang, and Chengxia Wang contributed equally to this work.
# These two authors are co-corresponding authors.
Corresponding authors:

1. Weishi Li, MD., Department of Orthopaedics, Peking University Third Hospital, 49 North Garden Road, Haidian District, Beijing, 100191, China (E-mail: puh3liweishi@163.com).
2. Lei Hu, Ph.D, School of Mechanical Engineering and Automation, Beihang University, 37 Xueyuan Road, Haidian District, Beijing, 100191, China (E-mail: Hulei9971@buaa.edu.cn)

Short running head: Automatic laminectomy cutting plane planning.

# Abstract

**Study Design**: Applied research

**Objective**: This study aims to use artificial intelligence to realize the automatic planning of laminectomy, and verify the method.

**Methods**: We propose a two-stage approach for automatic laminectomy cutting plane planning. The first stage was the identification of key points. 7 key points were manually marked on each CT image. The Spatial Pyramid Upsampling Network (SPU-Net) algorithm developed by us was used to accurately locate the 7 key points. In the second stage, based on the identification of key points, a personalized coordinate system was generated for each vertebra. Finally, the transverse and longitudinal cutting planes of laminectomy were generated under the coordinate system. The overall effect of planning was evaluated.

**Results**: In the first stage, the average localization error of the SPU-Net algorithm for the seven key points was $0.65\pm0.70$mm. In the second stage, a total of 320 transverse cutting planes and 640 longitudinal cutting planes were planned by the algorithm. Among them, the number of horizontal plane planning effects of grade A, B, and C were 318(99.38%), 1(0.31%), and 1(0.31%), respectively. The longitudinal planning effects of grade A, B, and C were 622(97.18%), 1(0.16%), and 17(2.66%), respectively.

**Conclusions**: In this study, we propose a method for automatic surgical path planning of laminectomy based on the localization of key points in CT images. The results showed that the method achieved satisfactory results. More studies are needed to confirm the reliability of this approach in the future.

## Background

Lumbar spinal stenosis (LSS) is a degenerative disease. As the growth of the age, degenerative changes in the intervertebral discs, ligamentum flavum and facet joints can lead to narrowing of the space in the spinal canal and compression of neurovascular structures. These changes can lead to pain in the back and lower limbs, mobility disorders as well as other disabilities[1]. In the United States, LSS is the most common reason for spinal surgery in patients over the age of 65[2]. Laminectomy is the standard treatment for symptomatic lumbar spinal stenosis refractory to conservative management[3]. Laminectomy is a delicate operation as some important neurovascular tissues around the lamina exist. Complications such as dural sac damage, nerve root damage, and cauda equina syndrome can occur if these tissues are damaged[4]. In order to minimize these injuries, surgeons should work in two ways: 1. Ensure that each cut does not penetrate the lamina to avoid damage from bone-cutting tools; 2. The decompression range of laminectomy should be reasonably planned to avoid the increased possibility of injury caused by repeated resection when the initial decompression range is too small. For the first point, the surgeon must maintain precise hand muscle control at all times, which is difficult during such a lengthy and difficult operation[5]. For the second point, it is necessary for the surgeon to be familiar with the anatomical structure of the lamina and be able to reasonably plan the decompression range of laminectomy through anatomical markers, which often require decades of experience. These factors contribute to a longer learning curve for young surgeons.

The emergence of the surgical robot provides a new way to solve this problem. At present, all the commercial spinal robots focus on the field of assisting pedicle screw placement, and no laminectomy-assisted robot is available[6]. Therefore, our team developed a collaborative spinal robot system for laminectomy[7]. This robot system with lamina penetrating recognition could reduce the damage due to operator fatigue. However, where the lamina should be cut cannot be determined by the system itself, requiring the surgeon to manually move the end of the bone cutting tool to the target point. To solve this problem, we need a method to plan the cutting plane of laminectomy. Qi et al.[8] reported a semi-automatic laminectomy cutting plane planning strategy that needs surgeons to set some professional instructions. However, this semi-automatic planning strategy has always relied on experienced surgeons, which will limit the clinical use of such robots. Therefore, the automatic laminectomy cutting plane planning (ALCPP) function is crucial. In addition to robot applications, ALCPP technology itself can also be combined with augmented reality and other technologies to provide visual reference for intraoperative operations and improve surgical safety.

In recent years, with the development of computer vision and artificial intelligence technology, the application of related technologies is also more and more wide. Tang et al. proposed a label-free vision method to detect the seismic performance of recycled aggregate CFST columns, which achieved good results [9]. Wang et al. proposed a long–close distance coordination control strategy for litchi picking that integrated target detection, point cloud clustering and instance segmentation algorithms [10]. Computer vision and artificial intelligence technology also play an important role in the field of medical image processing, such as the detection of spinal fractures 错误!未找到引用源。, the diagnosis of spinal arthritis [12], and the generation of 3D-printed surgical auxiliary instruments [13]. CT images play an extremely important role in the diagnosis of spinal diseases. Therefore, based on preoperative CT images, the use of computer vision and artificial intelligence technology has the potential to achieve ALCPP.

At present, there have been many studies focus on automatic planning of pedicle screw placement. Cai et al. achieved the positioning of two control points: the entry point and the direction indication point based on deep neural network, and realized the path planning[14]. Qi et al. established a local coordinate system through preoperative CT and realized the autonomous planning of surgical path by identifying feature points [15]. However, due to the relative complexity of laminectomy surgery, there are few studies on automatic laminectomy cutting plane planning. Qian et al. [16][17] used artificial intelligence to automatically extract the central point of the lamina from CT data, and calculated a reasonable laminectomy trajectory to achieve ALCPP. However, the lamina is not a regular structure, so the center point of the lamina is not a precise anatomical site. For this reason, errors are inevitable in the output of the lamina center point, as the author also mentioned in their study. In this study, we propose a new way to achieve ALCPP. Based on the experience of clinical experts, we selected 7 landmarks on the vertebra as the basis to build a personalized spatial coordinate system for each vertebra, and then generate the laminectomy plane.

Landmark detection is a research hotspot in the field of medical image analysis, and some researchers have achieved certain results through traditional machine learning methods. Lindner et al. realized hand joint localization in radiographs based on random forest regression voting [18]. Donner et al. achieved Global localization of 3D chemical structures based on pre-filtered Hough Forests and discrete optimization[19]. However, in recent years, convolutional neural network (CNN) has shown its powerful ability in the field including medical image processing, such as medical image segmentation [20-22], and multi-mode image registration [23-25]. In landmark detection task, CNN also showed better performance compared with traditional machine learning methods. At present, the localization of landmark based on CNN mainly focuses on the in-depth improvement of excellent algorithms such as U-Net [20], and the end-to-end generation of heatmap highlighting landmark [18,26]. In our method, based on the same idea, we developed a landmark detection algorithm called Spatial Pyramid Upsampling Network (SPU-Net) with multi-scale feature fusion ability based on the principle of heatmap regression. The purpose of this study is to clarify this method and verify its effectiveness.

## Materials and Methods

In order to generate the laminectomy cutting plane automatically according to the characteristics of each vertebra, we detected and located 7 landmarks. They are the center point A of the anterior edge of the vertebral body, the center point B of the posterior edge of the vertebral body, the medial edge point C and the lower edge point D of the left pedicle, the lower edge point E and the medial edge point F of the right pedicle, and the midpoint G on posterior side of the lower endplate (Figure.1). After the precise positioning of those 7 landmarks, a spatial coordinate fitting algorithm was used to generate a personalized spatial coordinate system for the vertebra. Finally, the laminectomy cutting plane is generated automatically in this coordinate system by combining the experience of surgeons and the location information of landmarks.

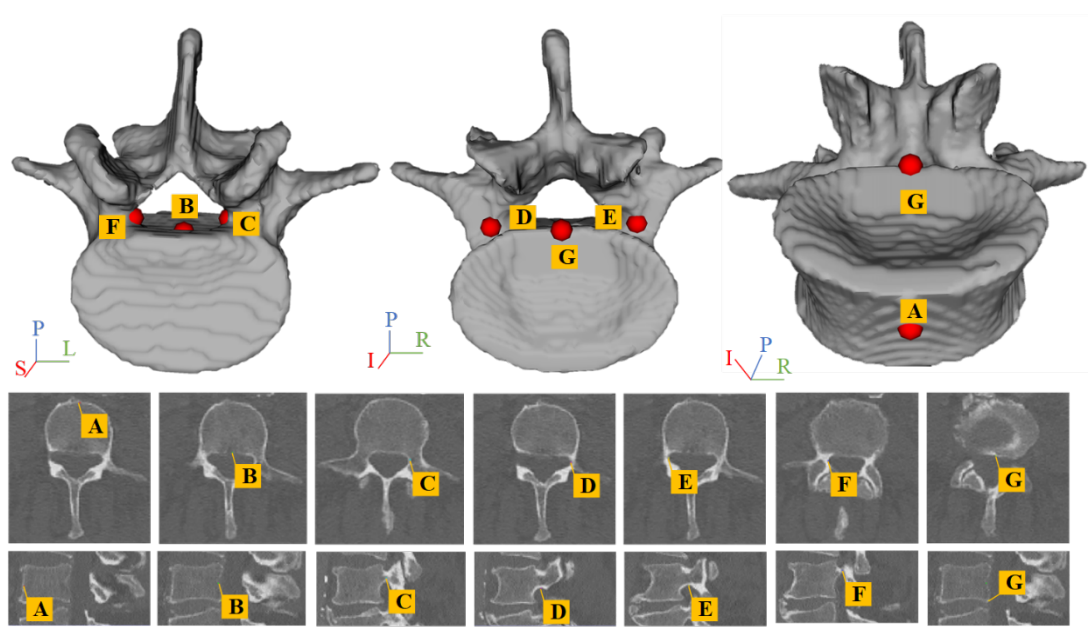

**Figure 1.** The selected landmarks. The coordinate system at the lower left corner of the 3-D vertebral model is the annotation of the spatial direction of the vertebra. P represents Posterior, S represents Superior, I represents Inferior, L represents Left, and R represents Right. The lower part of the figure shows the position of points A-G on the 2-D CT image.

We developed an effective landmark detection algorithm called SPU-Net to accurately locate these landmarks, which integrates multi-scale feature information and has good accuracy and robustness. The following is a detailed description of the related methods of SPU-Net and landmark-based laminectomy cutting plane generation.

## 2.1 SPU-Net

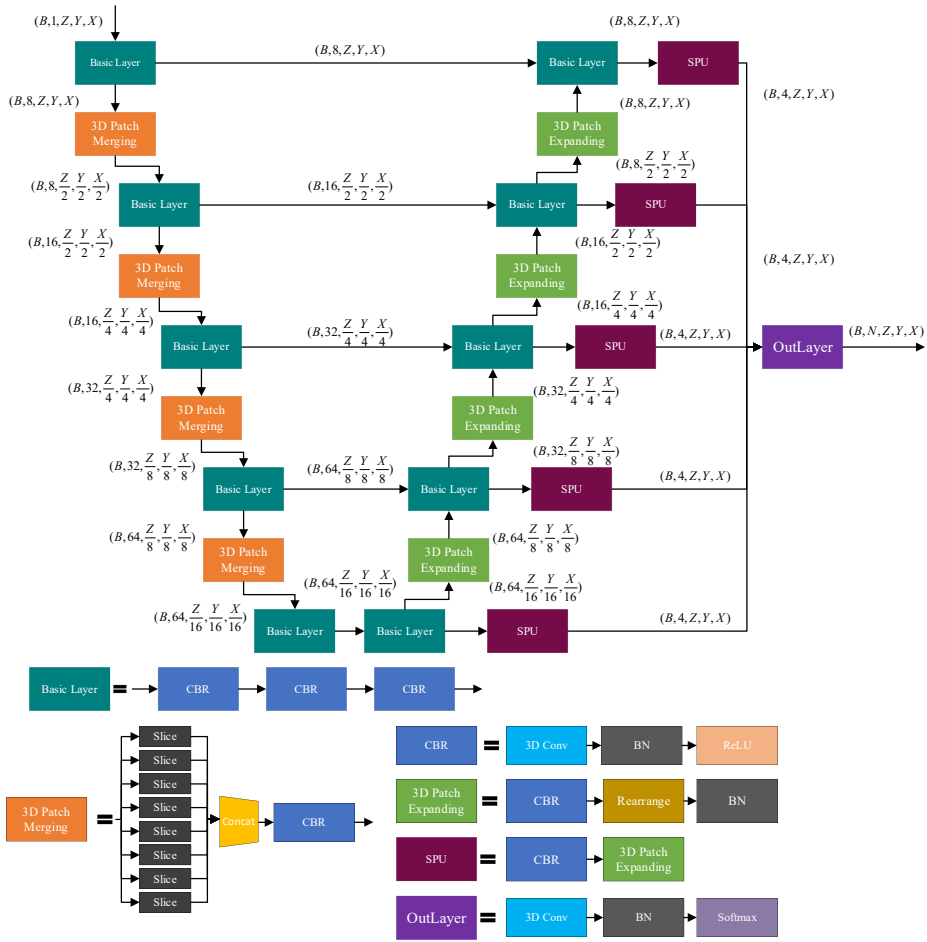

**Figure 2.** The structure chart of SPU-Net.

The SPU-Net we designed is deeply improved based on the basic architecture of U-Net[20]. SPU-Net also includes encoder and decoder. The specific modules in the network are described in Figure 2. Due to the excellent performance of Patch Merging and Patch Expanding in Swin-UNet[27], we used it in SPU-Net to replace pooling (which may lose information) and upsampling (which do not have the ability to learn). However, the original Patch Merging and Patch Expanding did not possess the processing capability of three-dimensional images, so we conducted three-dimensional expansion respectively according to their operating principles. The corresponding operation process is shown in figure 3.

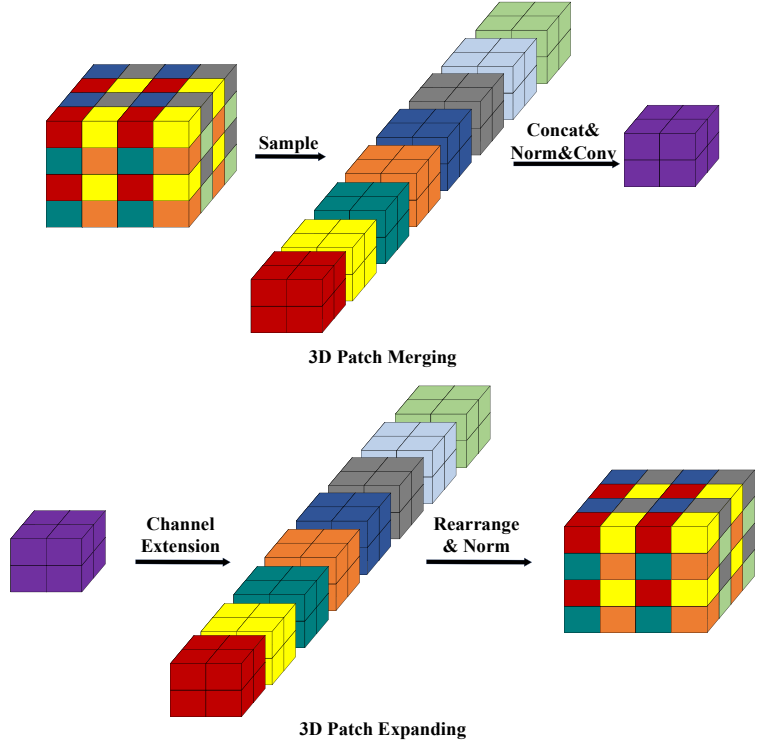

**Figure 3.** The operation process of 3D Patch Merging and 3D Patch Expanding.

3D Patch Merging is mainly used to implement downsampling to replace pooling operations. In the 3D Patch Merging, assuming that the shape of the original feature tensor is $(B,C,Z,Y,X)$, we obtained 8 feature tensors whose shape is $(B,C,\frac{Z}{2},\frac{Y}{2},\frac{X}{2})$ by interval sampling. Then, the feature tensor with shape of $(B,8C,\frac{Z}{2},\frac{Y}{2},\frac{X}{2})$ was obtained by splicing the feature channel dimension. Finally, the feature tensor with shape of $(B,C,\frac{Z}{2},\frac{Y}{2},\frac{X}{2})$ was obtained by processing it with three-dimensional convolution and BatchNorm. Therefore, the 3D Patch Merging operation in SPU-Net reduces the size of the input feature tensor by half, but does not affect the number of feature channels.

3D Patch Expanding is mainly used for upsampling to replace general upsampling operations. In SPU-Net, unlike 3D Patch Merging, the magnification of 3D Patch Expanding is configurable. In the 3D Patch Expanding, assuming that the shape of the original feature tensor is $(B,C,Z,Y,X)$, we can expand the number of feature channels through a 3D convolution operation. Suppose that we need to expand the size of the feature tensor to $S$ times of the input feature tensor, our feature channel needs to expand to $S^3$ times of the original. Therefore, a feature tensor with shape $(B,S^3 \times C,Z,Y,X)$ can be obtained by expanding the number of feature channels. On this basis, we rearrange them to obtain a feature tensor with shape $(B,C,S \times Z,S \times Y,S \times X)$. Therefore, the 3D Patch Expanding operation in SPU-Net will enlarge the size of the input feature tensor by a factor of $S$, but will not affect the number of feature channels.

In SPU-Net, we use the structure of Spatial pyramid upsampling to realize the size unification of multi-scale feature images, so as to realize the feature fusion of feature images. In the SPU structure, we first use a three-dimensional convolution operation to compress and unify the number of feature channels in each feature graph to reduce the subsequent calculation. After realizing the compression and unification of channel number of each feature image, we applied 3D Patch Expanding to achieve the unification of Shape. Therefore, the SPU structure can unify the feature images of each stage of the decoder part of SPU-Net into the same shape. On this basis, we spliced all feature images of uniform size in feature dimension and input them into OutLayer for processing to obtain thermal image $H$ which is the same as the number $N$ of target landmark. Finally, we only need to find the point with the highest brightness in each channel to achieve accurate positioning of corresponding landmarks. For example, for landmark $P_i, i = \{1,...,N\}$, we only need to search inside, and the positioning method is as follows:

$$P_i = argmax(H_i^p) \qquad (1)$$

**2.2 Cutting plane generation**

After using SPU-Net to locate the 7 landmarks of the vertebra, we developed a coordinate system fitting method to generate a personalized spatial coordinate system for that vertebra. Firstly, we take $\overrightarrow{AB}$ as the normal vector to establish a virtual plane passing through point B, and project all points C, D, E and F onto this plane, corresponding to points C', D', E' and F' respectively, the spatial relationship shown in Figure 4. After the projection is completed, the midpoints of C' and D', E' and F' are calculated as points H' and I' respectively. We set up a personalized spatial coordinate system with point B as the origin, $\overrightarrow{AB}$ as z-axis and $\overrightarrow{IH}$ as Y-axis.

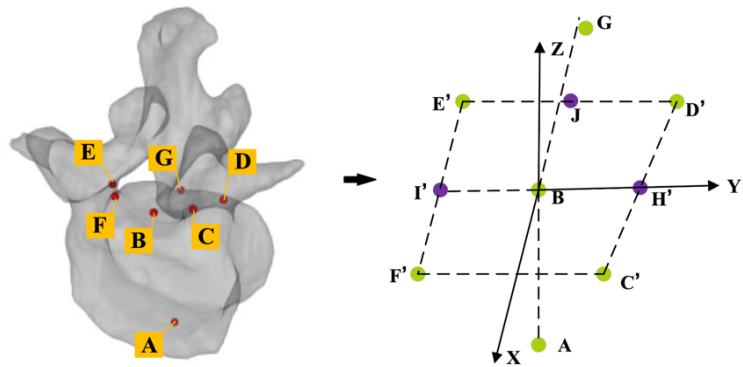

**Figure 4.** Personalized spatial coordinate system. the points C', D', E', and F' in the right part of the figure are the projections of the points C, D, E, and F onto the virtual plane which take $\overrightarrow{AB}$ as the normal vector and pass through point B.

Then, as shown in Figure 5, we planned three cutting planes for each vertebra according to clinical experience, all of which were parallel to the Z-axis. Point J is the midpoint of D'E', which means that point J doesn't have to be on the X-axis. We set the cutting plane 1 and 2 to be perpendicular to the Y-axis, and set plane 1 and plane 2 to be in the positive and negative direction of the Y-axis respectively, and pass-through point M and point N respectively. The distance between point M and point N to the coordinate origin in the Y-axis direction is 75% of the distance between point C 'and point F' to the coordinate origin in the Y-axis direction. For plane 3, we set it perpendicular to and in the negative direction of the X-axis, and it crosses point K, and the distance from point K to point J is 40% of the distance from point J to point G. In this way, personalized surgical planes can be generated.

For the total laminectomy task, we only need planes 1 and 2; For partial laminectomy, planes 1-3 are required.

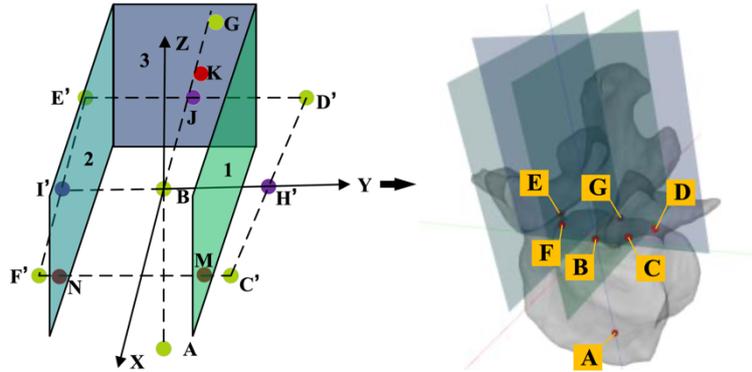

**Figure 5.** Laminectomy cutting plane generation. Point J, K, and G don't have to be on the X-axis

## 2.3 Experiment and training setup

We trained and tested SPU-Net under the Pytorch framework. In the construction of the SPU-Net dataset, 217 CT data from Verse2020 dataset[28][29][30] were used, to ensure the robustness of the algorithm as much as possible. This open-source dataset does not include private data, and there is no need to obtain informed consent from patients. Firstly, for each spinal segment, we used a personalized cuboid bounding box to intercept the CT image, so as to ensure that the cropped CT image only contained this complete spinal segment and contained as little information as possible about other spinal segments. In our previous work, we achieved accurate positioning of each spinal segment[31]. This is beneficial to reduce the interference caused by other adjacent segments to SPU-Net. The following figure shows the operation process.

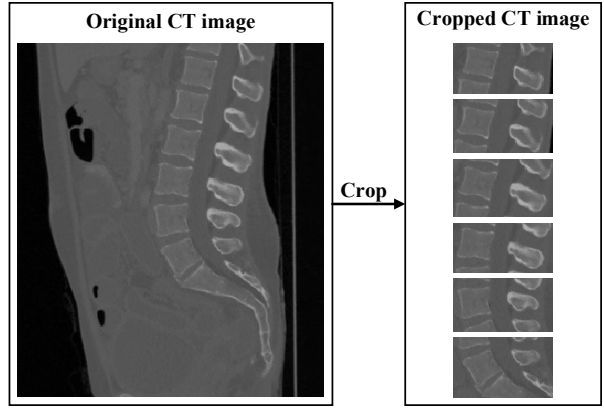

**Figure 6.** Schematic of CT image processing.

After the interception operation, 1302 CT images were generated, including 782 CT images in the training set, 260 CT images in the verification set and 260 CT images in the test set. The 7 landmarks were manually annotated by an attending physician with 2 years of experience to ensure the reliability of the data. We have developed a standard SOP process for those 7 landmarks labeling to ensure that it is repeatable. The details are as follows. Point A and B, the centre point of the anterior and posterior edge of the vertebral body: by constantly adjusting the sagittal plane and the transverse plane, the layer closest to the midline transverse plane of the vertebral body was selected. In this layer, the point at the anterior edge of the vertebral body was marked as point A, and the point at the posterior edge was marked as point B; Point C, the medial edge of the left pedicle: by constantly adjusting the sagittal plane and the transverse plane, the layer closest to the midline transverse plane of the left pedicle was selected. In this layer, the point at the medial edge of the left pedicle was marked as point C; The same for point F, the medial edge of the right pedicle; Point D, the lower edge of the left pedicle: by constantly adjusting the sagittal plane and the transverse plane, the layer closest to the midline sagittal plane of the left pedicle was selected. In this layer, the point at the lower edge of the left pedicle was marked as point D; The same for point E, the lower edge of the right pedicle; Point G, the midpoint on posterior side of the lower endplate: by constantly adjusting the sagittal plane and the transverse plane, the lowest transverse plane layer of the vertebral body was selected. The point at the posterior edge was marked as point G. After labeling, local CT images were scaled to a fixed size (72×128×128), and corresponding spatial key points were mapped to the set fixed size. We adjusted the window width and window position of the original local CT image to improve the contrast. The calculation is as follows.

$$P_{Des} = \begin{cases} 0 & P_{Src} \leq W_{min} \\ \dfrac{P_{Src} - W_{min}}{W_{max} - W_{min}} & W_{min} < P_{Src} < W_{max} \\ 1 & P_{Src} \geq W_{max} \end{cases} \quad (2)$$

In our method, we set $W_{min}$=-200 and $W_{max}$=600. For the landmark $P_i, i = \{1,...,N\}$, the corresponding target thermal diagram is generated in the following way:

$$H_i^t(x,\sigma) = \dfrac{1}{(2\pi)^{3/2}\sigma^3} \exp(-\dfrac{\|x - \dot{x}_i\|_2^2}{2\sigma^2}) \quad (3)$$

Where $H_i^t$ is the target thermal diagram of the ith channel and the coordinates of the target key points, $\dot{x}_i$ is the coordinate of the target key point. In the process of network training, we use MSELoss to supervise the network optimization training. In SPU-Net, the variation of feature tensor has been detailed in the network structure diagram. All convolution operations in the network are three-dimensional convolution, and the relevant parameters are set as $kernel\_size = (3,3,3)$, $stride = (1,1,1)$ and $padding = (1,1,1)$. The number of characteristic channels of input and output can be calculated according to the shape of the correlation tensor in the network structure diagram. The initial learning rate of network training was set at 1e$^{-3}$ and decreased by half every 40 rounds during training. In addition, network training is optimized by Adam optimizer. For landmark detection task, we used the Euclidean distance between the target point $L_t = (x_t, y_t, z_t)$ and the output point $L_p = (x_p, y_p, z_p)$ predicted of SPU-Net as the evaluation criterion to describe the positioning accuracy of SPU-NET. It is calculated as follows:

$$Dis = \sqrt{(x_p - x_t)^2 + (y_p - y_t)^2 + (z_p - z_t)^2} \quad (4)$$

**2.4 Evaluation of the effect of laminectomy cutting plane planning**

Due to the lack of evaluation system for the effect of ALCPP in the past, we developed a grading system according to the consensus of experts in our hospital (Figure. 7) as below:

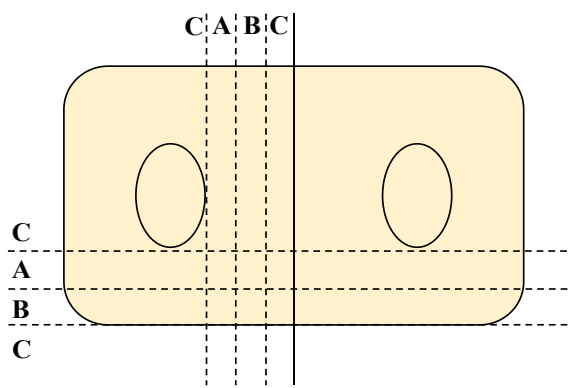

**Figure 7.** Grading system of the effect of laminectomy cutting plane planning.

Grade A, excellent: the cutting plane is basically perpendicular to the coronal plane of the vertebra; The longitudinal cutting plane is located between the midsagittal plane of the vertebra and the sagittal plane of medial edge of the pedicle, and was located in the lateral 1/3 region. The transverse cutting plane is located between the level of the lower edge of the pedicle and the level of the lower endplate of the vertebra, and is located in the cephalic half.

Grade B, good: the cutting plane is basically perpendicular to the coronal plane of the vertebra; The longitudinal cutting plane is located between the midsagittal plane of the vertebra and the sagittal plane of medial edge of the pedicle, and was located in the middle 1/3 region. The transverse cutting plane is located between the level of the lower edge of the pedicle and the level of the lower endplate of the vertebra, and is located in the caudal half.

Grade C, poor: The cutting plane is not perpendicular to the coronal plane of the vertebral body; The longitudinal cutting plane is located between the midsagittal plane of the vertebra and the sagittal plane of medial edge of the pedicle, and was located in the medial 1/3 region. Or the cutting plane is on the lateral side of the medial edge of the pedicle. Or the cutting plane overpasses the midsagittal plane to the opposite side. The transverse cutting plane is located above the level of the lower edge of the pedicle or below the level of the lower endplate of the vertebra.

Another 64 CT image data from Verse2020 dataset were used to test our cutting plane planning method. The planning effect of each cutting plane was evaluated according to the ALCPP grading system mentioned above.

**Results**

After 200 rounds of training, our SPU-Net algorithm finally achieved an average positioning error of 0.65±0.70mm, which was satisfactory. Compared with previous landmark detection algorithms based on convolutional neural networks, we substituted pooling and upsampling by 3D Patch Merging without losing information and 3D Patch Expanding with learning ability. In addition, SPU structure is adopted in our SPU-Net to achieve multi-scale feature fusion, which improves the performance of the algorithm. The detection effect of SPU-Net for landmark is shown in Figure 8.

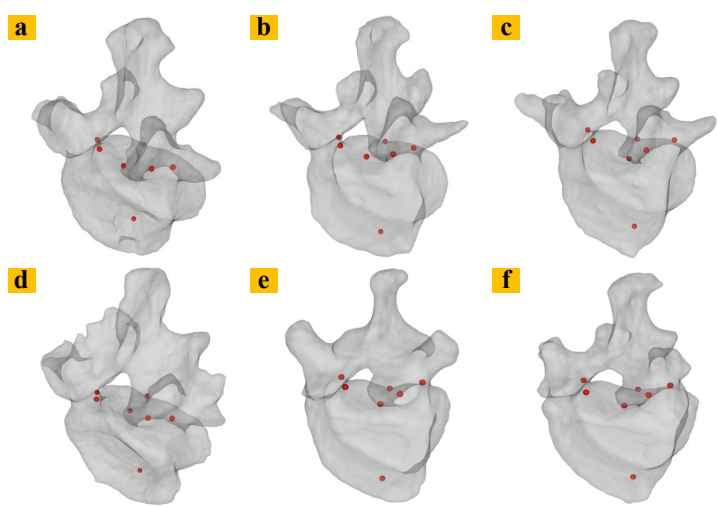

**Figure 8.** Schematic diagram of the detection effect of SPU-Net on Landmark

After the evaluation of ALCPP of 320 vertebrae in 64 spines of CT data, a total of 640 longitudinal cutting planes were planned, among which 622 (97.18%) were evaluated as grade A, 1 (0.16%) as grade B, and 17 (2.66%) as grade C. A total of 320 transverse cutting planes were planned, of which 318 (99.38%) were evaluated as grade A, 1 (0.31%) as grade B, and 1 (0.31%) as grade C. The result was satisfactory (Table 1, Figure 9).

Table 1. The results of the effect of automatic laminectomy cutting plane planning.

|  | Longitudinal cutting plane | Transverse cutting plane |
| --- | --- | --- |
| Grade A, excellent | 622 (97.18%) | 318 (99.38%) |
| Grade B, good | 1 (0.16%) | 1 (0.31%) |
| Grade C, poor | 17 (2.66%) | 1 (0.31%) |
| Total | 640 | 320 |

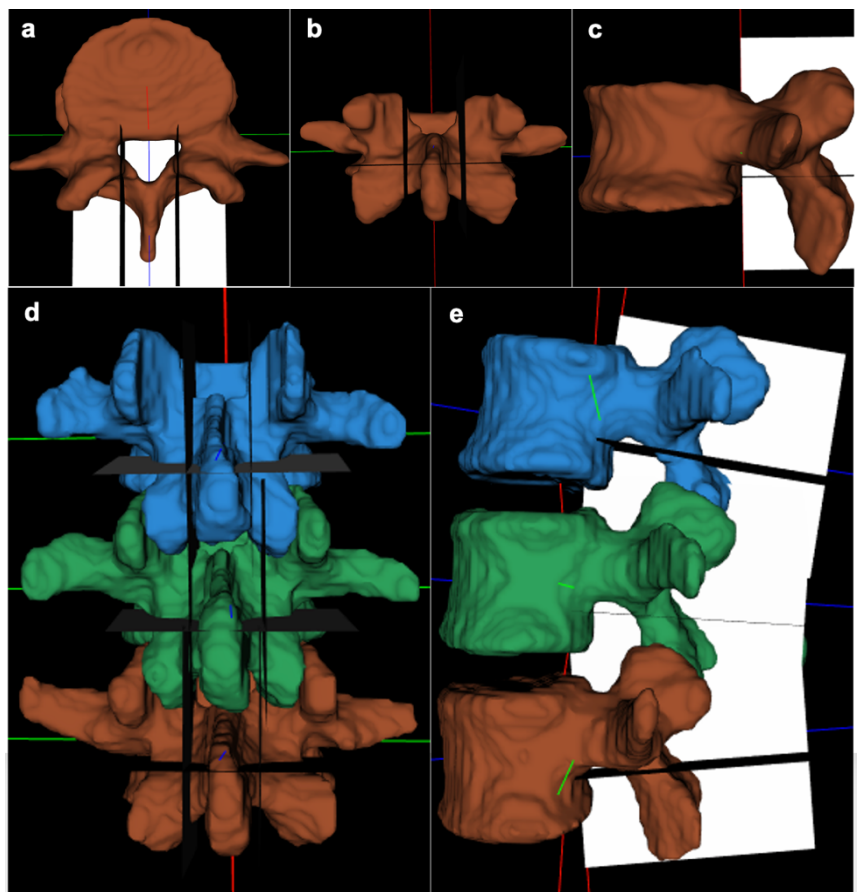

**Figure 9.** a-c, effect of single vertebra laminectomy cutting plane planning; d-e, effect of laminectomy cutting plane planning for multiple vertebrae.

## Discussion

The planning effect of our method has excellent results in most cases, but there are also a certain proportion of poor effects, as can be seen from Table 1. This is mainly because of the serious noise in some CT images and the severe deformity of some spines, which makes the location of some landmarks deviated greatly by SPU-Net, thus leading to the failure of planning. Therefore, in the actual robotic surgery planning system, it is necessary to retain the operation interface related to the surgeon's inspection and intervention, and manually correct it in the case of planning failure. In addition, we will continue to collect relevant case data to expand the dataset and improve the robustness of our method.

The original technique for laminectomy was proposed by Gill et al. [32], who suggested that the entire bony arch, including both inferior and adjacent superior articular processes, be completely removed. However, excessive resection can easily lead to iatrogenic instability of the spine and the resulting lumbar pain[33]. Grabias et al. [34] stated that in general "the medial half of the facet must be removed, and not the entire facet". The stability of the spine is usually ensured when the lateral half of the facet and interarticular remain. This is generally accepted. Even with posterior lumbar interbody fusion (PLIF) surgery, the decompression range requires partial preservation of the facet joint[35]. Therefore, the choice of the left and right boundaries of laminar decompression requires both sufficient decompression range and sufficient stability for the remainder of the spine. For the upper boundary of laminectomy, not all patients need complete laminectomy. Most of the stenosis occurs at the intervertebral level. Partial laminectomy can achieve the purpose of decompression, better protect the posterior ligament complex and reduce the occurrence of adjacent segment degeneration. The excision of the upper lamina in PLIF surgery is usually described as the lower 1/3 of the lamina[35]. It was found that the description of the boundary of laminectomy in the previous literature was based on the location of the lamina (e.g., resection of the medial or lower half of the lamina). However, the purpose of laminectomy is not to cut off the lamina, but to decompress the compressed spinal cord or nerve roots in the spinal canal by removing the lamina. In other words, the optimal planning of laminectomy should be based directly on the location of the spinal cord or nerve roots. According to this idea, our proposed ALCPP evaluation system directly uses the position of the pedicle and lower endplate rather than the position of the lamina for classification. That's because the nerve root is going down and out along the medial edge and the lower edge of the pedicle. When the longitudinal cutting plane (planes 1 and 2) is located at the lateral 1/3 region between the midsagittal plane of the vertebra and the sagittal plane of medial edge of the pedicle, the lateral nerve roots and the central dural sac could be fully exposed and decompressed, so we positioned this area as grade A, excellent; For the transverse cutting plane (plane 3), it needs to be above the level of the lower endplate of the upper vertebra, as stenosis is often located at the intervertebral level. At the same time, to ensure that the dura does not compress at the boundary after decompression or that the operative field is fully exposed in the case of intervertebral fusion, this plane should be located at some distance from the lower endplate. Finally, based on the clinical experience of our center, we rated the cephalic half area between the lower endplate and the lower edge of the pedicle as grade A, excellent.

Qian et al.[16] achieved automatic laminectomy planning by identifying the central point of the lamina in their study and reported the average positioning error was 2.37 mm. They stated that the lamina and its centre have been identified successfully in most cases, but there have been some failures. In fact, laminar segmentation is a challenging task because the lamina is irregular in shape. Therefore, based on the ALCPP evaluation system proposed before, we constructed an automatic laminectomy planning scheme from another perspective, avoiding lamina segmentation. By detecting and locating 7 landmarks including the medial and lower edges of the bilateral pedicles, we generated coordinates for each vertebra and generated the laminectomy plane according to grade A area. The results verify the feasibility of the method. In addition, because this method does not need to identify the shape of the lamina, it still has a good planning effect in the case of skewed lamina or spina bifida (Figure 10).

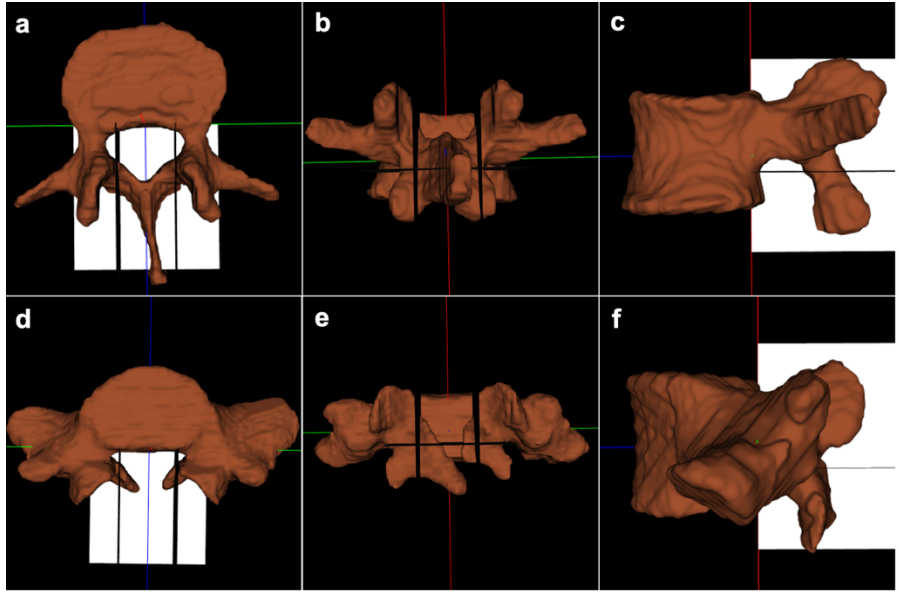

**Figure 10.** a-c, the effect of laminectomy cutting plane planning when spinous process and lamina deviate to one side; d-f, laminectomy cutting plane planning in the presence of spina bifida.

This study has some limitations. First of all, our data set is mainly composed of normal spinal CT images, and the proportion of severe spinal deformity images is low. As a result, our algorithm may have problems in landmarks detection of spinal severe deformity images, or even cause the failure of planning. Secondly, our algorithm has some limitations in processing CT image noise. When the image noise is very significant, planning failure may also occur. Finally, at present, the planning system is only in the laboratory stage, in the actual robotic surgery planning system, it is necessary to retain the operation interface related to the surgeon's inspection and intervention, and manually correct it in the case of planning failure. In the following research, we will not only improve the robustness of the algorithm by expanding data sets and random data enhancement, but also study image noise reduction and enhancement technology, so as to reduce the interference of noise and other adverse factors on landmark detection.

## Conclusions

In this study, we proposed an automatic cutting plane planning strategy for laminectomy based on landmark detection in CT images. The results show that the effect of this strategy is satisfactory. More studies are needed in the future to confirm the reliability of this strategy.

## List of abbreviations

| | |
|---|---|
| LSS | Lumbar spinal stenosis |
| ALCPP | Automatic laminectomy cutting plane planning |
| CNN | Convolutional neural network |
| SPU-Net | Spatial Pyramid Upsampling Network |
| PLIF | Posterior lumbar interbody fusion |

## Data Availability

The datasets used and/or analyzed during the current study are available from the corresponding author on reasonable request.

## Conflicts of Interest

The authors declare that there is no conflict of interest regarding the publication of this paper.

## Funding Statement

This work was supported by Natural Science Foundation of China (U20A20199), Beijing Natural Science Foundation (No. L202010) and National Key Research and Development Program of China (2018YFB1307604).

# Acknowledgments

None

# Supplementary Materials

None

**Figure legends**

Figure 1. The selected landmarks. The coordinate system at the lower left corner of the 3-D vertebral model is the annotation of the spatial direction of the vertebra. P represents Posterior, S represents Superior, I represents Inferior, L represents Left, and R represents Right. The lower part of the figure shows the position of points A-G on the 2-D CT image.

Figure 2. The structure chart of SPU-Net.

Figure 3. The operation process of 3D Patch Merging and 3D Patch Expanding.

Figure 4. Personalized spatial coordinate system. the points C', D', E', and F' in the right part of the figure are the projections of the points C, D, E, and F onto the virtual plane which take $\overrightarrow{AB}$ as the normal vector and pass through point B.

Figure 5. Laminectomy cutting plane generation. Point J, K, and G don't have to be on the X-axis.

Figure 6. Schematic of CT image processing.

Figure 7. Grading system of the effect of laminectomy cutting plane planning.

Figure 8. Schematic diagram of the detection effect of SPU-Net on Landmark.

Figure 9. a-c, effect of single vertebra laminectomy cutting plane planning; d-e, effect of laminectomy cutting plane planning for multiple vertebrae.

Figure 10. a-c, the effect of laminectomy cutting plane planning when spinous process and lamina deviate to one side; d-f, laminectomy cutting plane planning in the presence of spina bifida.

**Table legends**

Table 1. The results of the effect of automatic laminectomy cutting plane planning.